
\newbox\leftpage \newdimen\fullhsize \newdimen\hstitle \newdimen\hsbody
\tolerance=1000\hfuzz=2pt
\def\printertype{ps: }
\def\qms{\def\printertype{qms: }
\ifx\answ\bigans\else\voffset=-.4truein\hoffset=.125truein\fi}
\def\bigans{b }
%
\let\answ\bigans
\ifx\answ\bigans\message{This will come out unreduced.}
\magnification=1200\baselineskip=12pt plus 2pt minus 1pt
\hsbody=\hsize \hstitle=\hsize 
\else\message{(This will be reduced.} \let\lr=L
\magnification=1000\baselineskip=16pt plus 2pt minus 1pt
\voffset=-.31truein\vsize=7truein\hoffset=-.59truein
\hstitle=8truein\hsbody=4.75truein\fullhsize=10truein\hsize=\hsbody
\output={\ifnum\pageno=0 
  \shipout\vbox{\special{\printertype landscape}\makeheadline
    \hbox to \fullhsize{\hfill\pagebody\hfill}}\advancepageno
  \else
  \almostshipout{\leftline{\vbox{\pagebody\makefootline}}}\advancepageno
  \fi}
\def\almostshipout#1{\if L\lr \count1=1 \message{[\the\count0.\the\count1]}
      \global\setbox\leftpage=#1 \global\let\lr=R
  \else \count1=2
    \shipout\vbox{\special{\printertype landscape}
      \hbox to\fullhsize{\box\leftpage\hfil#1}}  \global\let\lr=L\fi}
\fi
%
\catcode`\@=11 
\newcount\yearltd\yearltd=\year\advance\yearltd by -1900

%
%

\def\draftmode{\message{ DRAFTMODE }\def\draftdate{{\rm preliminary draft:
\number\month/\number\day/\number\yearltd\ \ \hourmin}}%
\headline={\hfil\draftdate}\writelabels\baselineskip=20pt plus 2pt minus 2pt
 {\count255=\time\divide\count255 by 60 \xdef\hourmin{\number\count255}
  \multiply\count255 by-60\advance\count255 by\time
  \xdef\hourmin{\hourmin:\ifnum\count255<10 0\fi\the\count255}}}
\def\nolabels{\def\wrlabel##1{}\def\eqlabeL##1{}\def\reflabel##1{}}
\def\writelabels{\def\wrlabel##1{\leavevmode\vadjust{\rlap{\smash%
{\line{{\escapechar=` \hfill\rlap{\sevenrm\hskip.03in\string##1}}}}}}}%
\def\eqlabeL##1{{\escapechar-1\rlap{\sevenrm\hskip.05in\string##1}}}%
\def\reflabel##1{\noexpand\llap{\noexpand\sevenrm\string\string\string##1}}}
\nolabels
%
\global\newcount\secno \global\secno=0
\global\newcount\meqno \global\meqno=1
\def\newsec#1{\global\advance\secno by1\message{(\the\secno. #1)}
\global\subsecno=0\xdef\secsym{\the\secno.}\global\meqno=1
\noindent{\bf\the\secno. #1}\writetoca{{\secsym} {#1}}
\par\nobreak\medskip\nobreak}
\xdef\secsym{}
\global\newcount\subsecno \global\subsecno=0
\def\subsec#1{\global\advance\subsecno by1\message{(\secsym\the\subsecno. #1)}
\bigbreak\noindent{\it\secsym\the\subsecno. #1}\writetoca{\string\quad
{\secsym\the\subsecno.} {#1}}\par\nobreak\medskip\nobreak}
\def\appendix#1#2{\global\meqno=1\global\subsecno=0\xdef\secsym{\hbox{#1.}}
\bigbreak\bigskip\noindent{\bf Appendix #1. #2}\message{(#1. #2)}
\writetoca{Appendix {#1.} {#2}}\par\nobreak\medskip\nobreak}
%
%
\def\eqnn#1{\xdef #1{(\secsym\the\meqno)}\writedef{#1\leftbracket#1}%
\global\advance\meqno by1\wrlabel#1}
\def\eqna#1{\xdef #1##1{\hbox{$(\secsym\the\meqno##1)$}}
\writedef{#1\numbersign1\leftbracket#1{\numbersign1}}%
\global\advance\meqno by1\wrlabel{#1$\{\}$}}
\def\eqn#1#2{\xdef #1{(\secsym\the\meqno)}\writedef{#1\leftbracket#1}%
\global\advance\meqno by1$$#2\eqno#1\eqlabeL#1$$}
%
\newskip\footskip\footskip10pt plus 1pt minus 1pt 
\def\f@@t{\baselineskip\footskip\bgroup\aftergroup\@foot\let\next}
\setbox\strutbox=\hbox{\vrule height9.5pt depth4.5pt width0pt}
\global\newcount\ftno \global\ftno=0
\def\foot{\global\advance\ftno by1\footnote{$^{\the\ftno}$}}
%
\newwrite\ftfile
\def\footend{\def\foot{\global\advance\ftno by1\chardef\wfile=\ftfile
$^{\the\ftno}$\ifnum\ftno=1\immediate\openout\ftfile=foots.tmp\fi%
\immediate\write\ftfile{\noexpand\smallskip%
\noexpand\item{f\the\ftno:\ }\pctsign}\findarg}%
\def\footatend{\vfill\eject\immediate\closeout\ftfile{\parindent=20pt
\centerline{\bf Footnotes}\nobreak\bigskip\input foots.tmp }}}
\def\footatend{}
%
%
\global\newcount\refno \global\refno=1
\newwrite\rfile
\def\ref{$^{\the\refno}$\nref}
\def\nref#1{\xdef#1{$^{\the\refno}$}\xdef\rfn{\the\refno}
\writedef{#1\leftbracket#1}%
\ifnum\refno=1\immediate\openout\rfile=refs.tmp\fi%
\global\advance\refno by1\chardef\wfile=\rfile\immediate%
\write\rfile{\noexpand\item{{\rfn}.\
}\reflabel{#1\hskip.31in}\pctsign}\findarg}%
\def\findarg#1#{\begingroup\obeylines\newlinechar=`\^^M\pass@rg}%
{\obeylines\gdef\pass@rg#1{\writ@line\relax #1^^M\hbox{}^^M}%
\gdef\writ@line#1^^M{\expandafter\toks0\expandafter{\striprel@x #1}%
\edef\next{\the\toks0}\ifx\next\em@rk\let\next=\endgroup\else\ifx\next\empty%
\else\immediate\write\wfile{\the\toks0}\fi\let\next=\writ@line\fi\next\relax}}%
\def\striprel@x#1{} \def\em@rk{\hbox{}}

\def\addref#1{\immediate\write\rfile{\noexpand\item{}#1}} 
\def\startrefs#1{\immediate\openout\rfile=refs.tmp\refno=#1}
\def\xref{\expandafter\xr@f}\def\xr@f[#1]{#1}
\def\refs#1{[\r@fs #1{\hbox{}}]}
\def\r@fs#1{\edef\next{#1}\ifx\next\em@rk\def\next{}\else
\ifx\next#1\xref #1\else#1\fi\let\next=\r@fs\fi\next}
%

%
\newwrite\ffile\global\newcount\figno \global\figno=1
\def\fig{fig.~\the\figno\nfig}
\def\nfig#1{\xdef#1{fig.~\the\figno}%
\writedef{#1\leftbracket fig.\noexpand~\the\figno}%
\ifnum\figno=1\immediate\openout\ffile=figs.tmp\fi\chardef\wfile=\ffile%
\immediate\write\ffile{\noexpand\medskip\noexpand\item{Fig.\ \the\figno. }
\reflabel{#1\hskip.55in}\pctsign}\global\advance\figno by1\findarg}
\def\xfig{\expandafter\xf@g}\def\xf@g fig.\penalty\@M\ {}
\def\figs#1{figs.~\f@gs #1{\hbox{}}}
\def\f@gs#1{\edef\next{#1}\ifx\next\em@rk\def\next{}\else
\ifx\next#1\xfig #1\else#1\fi\let\next=\f@gs\fi\next}
\newwrite\lfile
{\escapechar-1\xdef\pctsign{\string\%}\xdef\leftbracket{\string\{}
\xdef\rightbracket{\string\}}\xdef\numbersign{\string\#}}

\def\writestop{\def\writestoppt{\immediate\write\lfile{\string\pageno%
\the\pageno\string\startrefs\leftbracket\the\refno\rightbracket%
\string\def\string\secsym\leftbracket\secsym\rightbracket%
\string\secno\the\secno\string\meqno\the\meqno}\immediate\closeout\lfile}}
\def\writestoppt{}\def\writedef#1{}
\def\seclab#1{\xdef #1{\the\secno}\writedef{#1\leftbracket#1}\wrlabel{#1=#1}}
\def\subseclab#1{\xdef #1{\secsym\the\subsecno}%
\writedef{#1\leftbracket#1}\wrlabel{#1=#1}}
\newwrite\tfile \def\writetoca#1{}
\def\leaderfill{\leaders\hbox to 1em{\hss.\hss}\hfill}
\def\writetoc{\immediate\openout\tfile=toc.tmp
   \def\writetoca##1{{\edef\next{\write\tfile{\noindent ##1
   \string\leaderfill {\noexpand\number\pageno} \par}}\next}}}

%
\ifx\answ\bigans
 
 \font\titlei=cmmi10 scaled\magstep3
\font\titleis=cmmi7 scaled\magstep3 \font\titleiss=cmmi5 scaled\magstep3
\font\titlesy=cmsy10 scaled\magstep3 \font\titlesys=cmsy7 scaled\magstep3
\font\titlesyss=cmsy5 scaled\magstep3 
\else
 
 \font\titlei=cmmi10 scaled\magstep4
\font\titleis=cmmi7 scaled\magstep4 \font\titleiss=cmmi5 scaled\magstep4
\font\titlesy=cmsy10 scaled\magstep4 \font\titlesys=cmsy7 scaled\magstep4
\font\titlesyss=cmsy5 scaled\magstep4 
\font\absrm=cmr10 scaled\magstep1 \font\absrms=cmr7 scaled\magstep1
\font\absrmss=cmr5 scaled\magstep1 \font\absi=cmmi10 scaled\magstep1
\font\absis=cmmi7 scaled\magstep1 \font\absiss=cmmi5 scaled\magstep1
\font\abssy=cmsy10 scaled\magstep1 \font\abssys=cmsy7 scaled\magstep1
\font\abssyss=cmsy5 scaled\magstep1 \font\absbf=cmbx10 scaled\magstep1
\skewchar\absi='177 \skewchar\absis='177 \skewchar\absiss='177
\skewchar\abssy='60 \skewchar\abssys='60 \skewchar\abssyss='60
\fi
\skewchar\titlei='177 \skewchar\titleis='177 \skewchar\titleiss='177
\skewchar\titlesy='60 \skewchar\titlesys='60 \skewchar\titlesyss='60
\ifx\answ\bigans\def\abstractfont{\footfont}\else
\def\abstractfont{\def\rm{\fam0\absrm}
\textfont0=\absrm \scriptfont0=\absrms \scriptscriptfont0=\absrmss
\textfont1=\absi \scriptfont1=\absis \scriptscriptfont1=\absiss
\textfont2=\abssy \scriptfont2=\abssys \scriptscriptfont2=\abssyss
\textfont\itfam=\bigit \def\it{\fam\itfam\bigit}
\textfont\bffam=\absbf \def\bf{\fam\bffam\absbf} \rm} \fi
\def\tenpoint{\def\rm{\fam0\tenrm}
\textfont0=\tenrm \scriptfont0=\sevenrm \scriptscriptfont0=\fiverm
\textfont1=\teni  \scriptfont1=\seveni  \scriptscriptfont1=\fivei
\textfont2=\tensy \scriptfont2=\sevensy \scriptscriptfont2=\fivesy
\textfont\itfam=\tenit \def\it{\fam\itfam\tenit}
\textfont\bffam=\tenbf \def\bf{\fam\bffam\tenbf} \rm}
%
%
\def\noblackbox{\overfullrule=0pt}
\hyphenation{anom-aly anom-alies coun-ter-term coun-ter-terms}
\def\inv{^{\raise.15ex\hbox{${\scriptscriptstyle -}$}\kern-.05em 1}}

\def\Dsl{\,\raise.15ex\hbox{/}\mkern-13.5mu D} 
\def\dsl{\raise.15ex\hbox{/}\kern-.57em\partial}
\def\del{\partial}

\font\bigit=cmti10 scaled \magstep1
\def\lspace{\ifx\answ\bigans{}\else\qquad\fi}
\def\lbspace{\ifx\answ\bigans{}\else\hskip-.2in\fi} 
\def\boxeqn#1{\vcenter{\vbox{\hrule\hbox{\vrule\kern3pt\vbox{\kern3pt
	\hbox{${\displaystyle #1}$}\kern3pt}\kern3pt\vrule}\hrule}}}
\def\mbox#1#2{\vcenter{\hrule \hbox{\vrule height#2in
		\kern#1in \vrule} \hrule}}  
%

\def\darr#1{\raise1.5ex\hbox{$\leftrightarrow$}\mkern-16.5mu #1}

\def\roughly#1{\raise.3ex\hbox{$#1$\kern-.75em\lower1ex\hbox{$\sim$}}}
\font\tenmss=cmss10
\font\absmss=cmss10 scaled\magstep1
\newfam\mssfam
\font\footrm=cmr8  \font\footrms=cmr5
\font\footrmss=cmr5   \font\footi=cmmi8
\font\footis=cmmi5   \font\footiss=cmmi5
\font\footsy=cmsy8   \font\footsys=cmsy5
\font\footsyss=cmsy5   \font\footbf=cmbx8
\font\footmss=cmss8
\def\footfont{\def\rm{\fam0\footrm}
\textfont0=\footrm \scriptfont0=\footrms
\scriptscriptfont0=\footrmss
\textfont1=\footi \scriptfont1=\footis
\scriptscriptfont1=\footiss
\textfont2=\footsy \scriptfont2=\footsys
\scriptscriptfont2=\footsyss
\textfont\itfam=\footi \def\it{\fam\itfam\footi}
\textfont\mssfam=\footmss \def\mss{\fam\mssfam\footmss}
\textfont\bffam=\footbf \def\bf{\fam\bffam\footbf} \rm}
\catcode`\@=12 
%
\newif\ifdraft

\noblackbox
\catcode`\@=11
\newif\iffrontpage
%
\ifx\answ\bigans
\def\titleft{\titsm}
\magnification=1200\baselineskip=12pt plus 2pt minus 1pt
%
\voffset=0.35truein\hoffset=0.250truein
\hsize=6.0truein\vsize=8.5 truein
\hsbody=\hsize\hstitle=\hsize
\else\let\lr=L
\def\titleft{\titla}
\magnification=1000\baselineskip=14pt plus 2pt minus 1pt
%
\vsize=6.5truein
\hstitle=8truein\hsbody=4.75truein
\fullhsize=10truein\hsize=\hsbody
\fi
\parskip=4pt plus 15pt minus 1pt
\font\titsm=cmr10 scaled\magstep2
\font\titla=cmr10 scaled\magstep3
\font\tenmss=cmss10
\font\absmss=cmss10 scaled\magstep1
\newfam\mssfam
\font\footrm=cmr8  \font\footrms=cmr5
\font\footrmss=cmr5   \font\footi=cmmi8
\font\footis=cmmi5   \font\footiss=cmmi5
\font\footsy=cmsy8   \font\footsys=cmsy5
\font\footsyss=cmsy5   \font\footbf=cmbx8
\font\footmss=cmss8
\def\footfont{\def\rm{\fam0\footrm}
\textfont0=\footrm \scriptfont0=\footrms
\scriptscriptfont0=\footrmss
\textfont1=\footi \scriptfont1=\footis
\scriptscriptfont1=\footiss
\textfont2=\footsy \scriptfont2=\footsys
\scriptscriptfont2=\footsyss
\textfont\itfam=\footi \def\it{\fam\itfam\footi}
\textfont\mssfam=\footmss \def\mss{\fam\mssfam\footmss}
\textfont\bffam=\footbf \def\bf{\fam\bffam\footbf} \rm}
\def\tenpoint{\def\rm{\fam0\tenrm}
\textfont0=\tenrm \scriptfont0=\sevenrm
\scriptscriptfont0=\fiverm
\textfont1=\teni  \scriptfont1=\seveni
\scriptscriptfont1=\fivei
\textfont2=\tensy \scriptfont2=\sevensy
\scriptscriptfont2=\fivesy
\textfont\itfam=\tenit \def\it{\fam\itfam\tenit}
\textfont\mssfam=\tenmss \def\mss{\fam\mssfam\tenmss}
\textfont\bffam=\tenbf \def\bf{\fam\bffam\tenbf} \rm}
\ifx\answ\bigans\def\abstractfont{\tenpoint}\else
\def\abstractfont{\def\rm{\fam0\absrm}
\textfont0=\absrm \scriptfont0=\absrms
\scriptscriptfont0=\absrmss
\textfont1=\absi \scriptfont1=\absis
\scriptscriptfont1=\absiss
\textfont2=\abssy \scriptfont2=\abssys
\scriptscriptfont2=\abssyss
\textfont\itfam=\bigit \def\it{\fam\itfam\bigit}
\textfont\mssfam=\absmss \def\mss{\fam\mssfam\absmss}
\textfont\bffam=\absbf \def\bf{\fam\bffam\absbf}\rm}\fi
%
\def\f@@t{\baselineskip10pt\lineskip0pt\lineskiplimit0pt
\bgroup\aftergroup\@foot\let\next}
\setbox\strutbox=\hbox{\vrule height 8.pt depth 3.5pt width\z@}
\def\vfootnote#1{\insert\footins\bgroup
\baselineskip10pt\footfont
\interlinepenalty=\interfootnotelinepenalty
\floatingpenalty=20000
\splittopskip=\ht\strutbox \boxmaxdepth=\dp\strutbox
\leftskip=24pt \rightskip=\z@skip
\parindent=12pt \parfillskip=0pt plus 1fil
\spaceskip=\z@skip \xspaceskip=\z@skip
\Textindent{$#1$}\footstrut\futurelet\next\fo@t}
\def\Textindent#1{\noindent\llap{#1\enspace}\ignorespaces}
\def\footnote#1{\attach{#1}\vfootnote{#1}}%

\def\foot{\attach\footsymbolgen\vfootnote{\footsymbol}}
\let\footsymbol=\star
\newcount\lastf@@t           \lastf@@t=-1
\newcount\footsymbolcount    \footsymbolcount=0
\def\footsymbolgen{\relax\footsym
\global\lastf@@t=\pageno\footsymbol}
\def\footsym{\ifnum\footsymbolcount<0
\global\footsymbolcount=0\fi
{\iffrontpage \else \advance\lastf@@t by 1 \fi
\ifnum\lastf@@t<\pageno \global\footsymbolcount=0
\else \global\advance\footsymbolcount by 1 \fi }
\ifcase\footsymbolcount \fd@f\star\or
\fd@f\dagger\or \fd@f\ast\or
\fd@f\ddagger\or \fd@f\natural\or
\fd@f\diamond\or \fd@f\bullet\or
\fd@f\nabla\else \fd@f\dagger
\global\footsymbolcount=0 \fi }
\def\fd@f#1{\xdef\footsymbol{#1}}
\def\space@ver#1{\let\@sf=\empty \ifmmode #1\else \ifhmode
\edef\@sf{\spacefactor=\the\spacefactor}
\unskip${}#1$\relax\fi\fi}
\def\attach#1{\space@ver{\strut^{\mkern 2mu #1}}\@sf}
%
\newif\ifnref
\def\rrr#1#2{\relax\ifnref\nref#1{#2}\else\ref#1{#2}\fi}
\def\ldf#1#2{\begingroup\obeylines
\gdef#1{\rrr{#1}{#2}}\endgroup\unskip}

\nreffalse
\def\refout{\footatend\immediate\closeout\rfile\writestoppt
\baselineskip=10pt\newsec{References}{\frenchspacing%
\parindent=12pt\escapechar=` \input refs.tmp\vfill\eject}\nonfrenchspacing}
%
\def\eqn#1{\xdef #1{(\secsym\the\meqno)}
\writedef{#1\leftbracket#1}%
\global\advance\meqno by1\eqno#1\eqlabeL#1}
\def\eqnalign#1{\xdef #1{(\secsym\the\meqno)}
\writedef{#1\leftbracket#1}%
\global\advance\meqno by1#1\eqlabeL{#1}}
%
\def\chap#1{\newsec{#1}}
\def\chapter#1{\chap{#1}}
\def\sect#1{\subsec{{ #1}}}
\def\section#1{\sect{#1}}
\def\\{\ifnum\lastpenalty=-10000\relax
\else\hfil\penalty-10000\fi\ignorespaces}
\def\note#1{\leavevmode%
\edef\@@marginsf{\spacefactor=\the\spacefactor\relax}%
\ifdraft\strut\vadjust{%
\hbox to0pt{\hskip\hsize%
\ifx\answ\bigans\hskip.1in\else\hskip .1in\fi%
\vbox to0pt{\vskip-\dp
\strutbox\sevenbf\baselineskip=8pt plus 1pt minus 1pt%
\ifx\answ\bigans\hsize=.7in\else\hsize=.35in\fi%
\tolerance=5000 \hbadness=5000%
\leftskip=0pt \rightskip=0pt \everypar={}%
\raggedright\parskip=0pt \parindent=0pt%
\vskip-\ht\strutbox\noindent\strut#1\par%
\vss}\hss}}\fi\@@marginsf\kern-.01cm}
\def\titlepage{%
\frontpagetrue\nopagenumbers\abstractfont%
\hsize=\hstitle\rightline{\vbox{\baselineskip=10pt%
{\abstractfont\pubnum}}}\pageno=0}
\frontpagefalse
\def\pubnum{}
\def\pdate{\number\month/\number\yearltd}
\def\makefootline{\iffrontpage\vskip .27truein
\line{\the\footline}
\vskip -.1truein\leftline{\vbox{\baselineskip=10pt%
{\abstractfont\pdate}}}
\else\vskip.5cm\line{\hss \tenrm $-$ \folio\ $-$ \hss}\fi}
\def\title#1{\vskip .7truecm\titlestyle{\titleft #1}}
\def\titlestyle#1{\par\begingroup \interlinepenalty=9999
\leftskip=0.02\hsize plus 0.23\hsize minus 0.02\hsize
\rightskip=\leftskip \parfillskip=0pt
\hyphenpenalty=9000 \exhyphenpenalty=9000
\tolerance=9999 \pretolerance=9000
\spaceskip=0.333em \xspaceskip=0.5em
\noindent #1\par\endgroup }
\def\autskip{\ifx\answ\bigans\vskip.5truecm\else\vskip.1cm\fi}
\def\author#1{\vskip .7in \centerline{#1}}

\def\address#1{\ifx\answ\bigans\vskip.2truecm
\else\vskip.1cm\fi{\it \centerline{#1}}}
\def\abstract#1{
\vskip .5in\vfil\centerline
{\bf Abstract}\penalty1000
{{\smallskip\ifx\answ\bigans\leftskip 2pc \rightskip 2pc
\else\leftskip 5pc \rightskip 5pc\fi
\noindent\abstractfont \baselineskip=12pt
{#1} \smallskip}}
\penalty-1000}
%

%


\def\bfone{\relax{\rm 1\kern-.35em 1}}
\def\inbar{\vrule height1.5ex width.4pt depth0pt}
\def\IC{\relax\,\hbox{$\inbar\kern-.3em{\mss C}$}}
\def\ID{\relax{\rm I\kern-.18em D}}
\def\IF{\relax{\rm I\kern-.18em F}}
\def\IH{\relax{\rm I\kern-.18em H}}
\def\II{\relax{\rm I\kern-.17em I}}
\def\IN{\relax{\rm I\kern-.18em N}}
\def\IP{\relax{\rm I\kern-.18em P}}
\def\IQ{\relax\,\hbox{$\inbar\kern-.3em{\rm Q}$}}
\def\IR{\relax{\rm I\kern-.18em R}}
\font\cmss=cmss10 \font\cmsss=cmss10 at 7pt
\def\ZZ{\relax\ifmmode\mathchoice
{\hbox{\cmss Z\kern-.4em Z}}{\hbox{\cmss Z\kern-.4em Z}}
{\lower.9pt\hbox{\cmsss Z\kern-.4em Z}}
{\lower1.2pt\hbox{\cmsss Z\kern-.4em Z}}\else{\cmss Z\kern-.4em Z}\fi}
\def\CP{\relax\ifmmode\mathchoice
{\hbox{\cmss CP}}{\hbox{\cmss CP}}
{\lower.9pt\hbox{\cmsss CP}}
{\lower1.2pt\hbox{\cmsss CP}}\else{\cmss CP}\fi}
\def\a{\alpha} \def\b{\beta} 
 
 \def\l{\lambda}
 
\def\cA{{\cal A}}

\def\cJ{{\cal J}} 
 \def\cM{{\cal M}}
 
 \def\cQ{{\cal Q}}
\def\cR{{\cal R}} 
\def\nup#1({{\it Nucl.\ Phys.}\ $\us {B#1}$\ (}
\def\plt#1({{\it Phys.\ Lett.}\ $\us  {#1}$\ (}
\def\cmp#1({{\it Comm.\ Math.\ Phys.}\ $\us  {#1}$\ (}
\def\prp#1({{\it Phys.\ Rep.}\ $\us  {#1}$\ (}
\def\prl#1({{\it Phys.\ Rev.\ Lett.}\ $\us  {#1}$\ (}
\def\prv#1({{\it Phys.\ Rev.}\ $\us  {#1}$\ (}
\def\mpl#1({{\it Mod.\ Phys.\ Let.}\ $\us  {A#1}$\ (}
\def\ijmp#1({{\it Int.\ J.\ Mod.\ Phys.}\ $\us{A#1}$\ (}
\def\tit#1|{{\it #1},\ }
%

%

\def\tilde{\widetilde}

\def\us#1{\bf{#1}}

\def\hyp{{\vrule height 1.9pt width 3.5pt depth -1.5pt}\hskip2.0pt}

\def\Coe#1.#2.{{#1\over #2}}
\def\coeff#1#2{\relax{\textstyle {#1 \over #2}}\displaystyle}
\def\coe#1.#2.{\relax{\textstyle {#1 \over #2}}\displaystyle}

\def\shalf{\relax{\textstyle {1 \over 2}}\displaystyle}

\def\to{\rightarrow}
\def\notin{\hbox{{$\in$}\kern-.51em\hbox{/}}}
\def\shdot{\!\cdot\!}

\def\attac#1{\Bigl\vert
{\phantom{X}\atop{{\rm\scriptstyle #1}}\phantom{X}}}

\def\del{\partial}

\def\nex#1{$N\!=\!#1$}

\def\cc{$^,$}

\catcode`\@=12
\def\nul#1,{{\it #1},}
\ldf\mat{D.J.~Gross and A.A.~Migdal, \prl{64} (1990) 717;
M.~Douglas and S.~Shenker, \nup{235} (1990) 635;
E.~Brezin and V.~Kazakov, \plt{236B} (1990) 144.}
\ldf\TOPALG{E.\ Witten, \cmp{117} (1988) 353; \cmp{118} (1988) 411;
\nup340 (1990) 281.}
\ldf\EYtop{T.\ Eguchi and S.\ Yang, \mpl4 (1990) 1693.}
\ldf\Muss{G.\ Mussardo, G.\ Sotkov, M.\ Stanishkov, {\it Int.\ J.\ Mod.\ Phys.}
{\us A4} (1989) 1135; N.\ Ohta and H.\ Suzuki, \nup332(1990) 146.}
\ldf\fusions{M.\ Spiegelglas and S.\ Yankielowicz, \nul{ G/G topological field
theories by cosetting G(k)}, preprint TECHNION-PH-90-34; D.\ Gepner, \cmp141
(1991) 381; M.\ Spiegelglas, \plt274(1992) 21.}
\ldf\MD{M.\ Douglas, \plt238B(1990) 176.}
\ldf\topgr{E.\ Witten,  \nup340 (1990) 281; E.\ and H.\ Verlinde, \nup348
(1991) 457.}
\ldf\blackh{E.\ Witten,   {\it Phys.\ Rev.} {\us D44} (1991) 314.}
\ldf\TFTmat{R.\ Dijkgraaf and E.\ Witten, \nup342(1990) 486;
R.\ Dijkgraaf and E.\ and H.\ Verlinde, \nup348 (1991) 435;
For a review, see: R.\ Dijkgraaf, \nul{ Intersection theory, integrable
hierarchies and topological field theory}, preprint IASSNS-HEP-91/91.}
\ldf\Witgr{E.\ Witten, \nup373 (1992) 187. }
\ldf\integr{P.\ Fendley, W.\ Lerche, S.\ Mathur and N.P.\ Warner, \nup348
(1991) 66; S.\ Cecotti and C.\ Vafa, \nup367(1991)359; D.\ Nemeschansky and
N.P.\ Warner, \nup 380(1992) 241.}
\ldf\wbrs{M.\ Bershadsky, W.\ Lerche, D.\ Nemeschansky and N.P.\ Warner, \plt
B292 (1992) 35; E.\ Bergshoeff, A.\ Sevrin and X.\ Shen, \nul{ A Derivation of
the BRST operator for noncritical W strings}, preprint UG-8-92; J. de Boer and
J. Goeree, \nul{ KPZ Analysis for $W_3$ gravity}, preprint THU-92/34.}
\ldf\DVV{R.\ Dijkgraaf, E. Verlinde and H. Verlinde, \nup{352} (1991) 59.}
\ldf\Loss{A. Lossev, \nul{ Descendants constructed from matter field and K.
Saito higher residue pairing in Landau-Ginzburg theories coupled to topological
gravity}, preprint TPI-MINN-92-40-T. }
\ldf\BGS{B.\ Gato-Rivera and A.M.\ Semikhatov, \plt B293 (1992) 72.}
\ldf\LVW{W.\ Lerche, C.\ Vafa and N.P.\ Warner, \nup324 (1989) 427.}
\ldf\cring{D.\ Gepner, \nul{ A comment on the chiral algebras of quotient
superconformal field theories}, preprint PUPT-1130; S.\ Hosono and A.\
Tsuchiya, \cmp136(1991) 451.}
\ldf\EXTRA{B.\ Lian and G.\ Zuckerman, \plt254B (1991) 417; P.\ Bouwknegt, J.\
McCarthy and K.\ Pilch, \cmp145(1992) 541; A.\ Polyakov, \mpl6(1991) 635; S.\
Mukherji, S.\ Mukhi and A.\ Sen, \plt 266B (1991) 337; E.\ Witten, \nup373
(1992) 187; H.\ Kanno and M.\ Sarmadi, preprint IC/92/150; K.\ Itoh and N.\
Ohta, \nup377 (1992) 113; N.\ Chair, V.\ Dobrev and H.\ Kanno, \plt283B (1992)
194; S.\ Govindarajan, T.\ Jayamaran, V.\ John and P.\ Majumdar, \mpl7 (1992)
1063; S.\ Govindarajan, T.\ Jayamaran and V.\ John, \nul{ Chiral rings and
physical states in $c<1$ string theory}, preprint IMSc-92/30.}
\ldf\GG{E.\ Witten, \nup371(1992)191; O.\ Aharony, O.\ Ganor, J.\ Sonnenschein,
S.\ Yankielowicz and N.\ Sochen, preprint TAUP-1961-92; \nul{ c=1 String theory
as a topological G/G model}, preprint TAUP-2032-93; O.\ Aharony, J.\
Sonnenschein and S.\ Yankielowicz, \plt B289 (1992) 309; J.\ Sonnenschein,
\nul{Physical states in topological coset models}, preprint TAUP-1999-92; V.\
Sadov, \nul{On the spectra of $sl(N)_k/sl(N)_k$ cosets and $W_N$ gravities I},
preprint HUTP-92/A055.}
\ldf\BMPtop{P.\ Bouwknegt, J.\ McCarthy and K.\ Pilch, \nul{ On physical states
in 2d (topological) gravity}, preprint CERN-TH.6645/92.}
\ldf\Wchiral{K.\ Ito, \plt B259(1991) 73; \nup370(1992) 123; D.\ Nemeschansky
and S.\ Yankielowicz, \nul{ N=2 W-algebras,
Kazama-Suzuki models and Drinfeld-Sokolov reduction}, preprint USC-91-005A;
L.J.\ Romans, \nup369 (1992) 403; W.\ Lerche, D.\ Nemeschansky and N.P.\
Warner, unpublished.}
\ldf\bo{M.\ Bershadsky and H.\ Ooguri, \cmp126(1989) 49;  M.~Bershadsky and
H.~Ooguri, \plt{229B} (1989) 374.}
\ldf\topw{K.\ Li, \plt B251 (1990) 54, \nup346 (1990) 329; H.\ Lu, C.N.\ Pope
and X.\ Shen, \nup366(1991) 95; S.\ Hosono, \nul{
Algebraic definition of topological W-gravity}, preprint UT-588.}
\ldf\DK{J.\ Distler and T.\ Kawai, \nup321(1989) 509.}
\ldf\MS{P.\ Mansfield and B.\ Spence, \nup362(1991) 294.}
\ldf\KS{Y.\ Kazama and H.\ Suzuki, \nup321(1989) 232.}
\ldf\Keke{K.\ Li, \nup354(1991) 711; \nup354(1991)725.}
\ldf\Vafa{C.\ Vafa, \mpl6 (1991) 337.}
\ldf\Ind{S.\ Govindarajan, T.\ Jayamaran and V.\ John, in ref.\EXTRA}
\ldf\modul{D.\ Kutasov, E.\ Martinec and N.\ Seiberg, \plt B276 (1992) 437.}
\ldf\dis{J.\ Distler, \nup342(1990) 523.}
\ldf\Wstrings{A.\ Bilal and J.\ Gervais, \nup326(1989) 222; P.\ Mansfield and
B.\ Spence, \nup362(1991) 294; S. Das, A. Dhar and S. Kalyana Rama, preprint
TIFR/TH/91-20; C.N.\ Pope, L.J.\ Romans and K.S.\ Stelle, \plt268B(1991) 167;
H.\ Lu, C.N.\ Pope, S.\ Schrans and K.\ Xu, \nul{ The complete Spectrum of the
$W$-String}, preprint CTP-TAMU-5/92, KUL-TF-92/1, KUL-TF-92/1; C.N.\ Pope,
\nul{A Review of W-strings}, preprint CTP-TAMU-30/92.}
\ldf\popextra{For extra states at $c_M=100$, see: C.N.\ Pope, E. Sezgin\ , K.S.
Stelle\ and X.J.\ Wang, \nul{Discrete states in the $W_3$ string}, preprint
CTP-TAMU-64-92; H. Lu, B.E.W. Nilsson, C.N. Pope, K.S. Stelle, P.C. West,
\nul{The low level spectrum of the $W_3$ string}, preprint CTP-TAMU-70-92.}
\ldf\BLNWB{M.\ Bershadsky, W.\ Lerche, D.\ Nemeschansky and N.P.\ Warner,
{\it Extended N=2 superconformal structure of gravity and W-gravity coupled to
matter}, preprint CALT-68-1832, CERN-TH.6694/92, HUTP-A061/92 and USC-92/021,
to appear in {\it Nucl.\ Phys.\ B}.}
\ldf\EYQ{T.\ Eguchi, H.\ Kanno, Y.\ Yamada and S.-K.\ Yang, \nul{ Topological
strings, flat coordinates and gravitational descendants}, preprint UT-630.}
\ldf\newBMP{P.\ Bouwknegt, J.\ McCarthy and K.\ Pilch, \nul{ Semi-infinite
cohomology of $W$-algebras}, preprint USC-93/11 and ADP-23-200/M15.}
\ldf\gep{D.\ Gepner, \nul{ Foundations of rational quantum field theory 1},
preprint CALT-68-1825.\goodbreak}
\ldf\MuVa{S.\ Mukhi and C.\ Vafa, \nul{ Two dimensional black hole as a
topological coset model of $c=1$ string theory}, preprint HUTP-93/A002,
TIFR/TH/93-01.}
\ldf\disp{I.\ Krichever, \cmp143 (1992) 415; B.\ Dubrovin, \nul{ Integrable
systems and classification of 2-dimensional topological field theories},
SISSA preprint 162/92.}
\ldf\Wrev{For reviews, see: P.\ Bouwknegt and K.\ Schoutens, \nul{ W-symmetry
in conformal field theory}, preprint CERN-TH.6583/92 and ITP-SB-92-23; C.M.
Hull, \nul{ Lectures on W-gravity, W-geometry and W-strings}, preprint
QMW-93-2.}
\def\brs{$BRST$\ }
\def\qbrs{\cQ_{BRST}}
\def\jbrs{\cJ_{BRST}}
\def\mm#1#2{\cM_{#1,#2}}

\def\bi#1{b_{#1}}
\def\ci#1{c_{#1}}
\def\zw#1#2.{{#2\over(z-w)^{#1}}}
\def\tg{T_{gh}}
\def\cQ{{\cal Q}}
\def\cW{{\cal W}}
\def\lv{{Liouville}}
\def\wnpq#1#2#3{\cW^{(#1)}_{#2,#3}}

\def\sc{superconformal\ }

\def\qt{\tilde\cQ}
\def\zmw#1{(z-w)^{#1}}

\def\cW{{\cal W}}
\def\g{{\gamma^0}}

\def\LG{Lan\-dau-Ginz\-burg\ }

\def\RG{\cR^{{\rm gr}}}
\def\RN{\cR^{\Atop}}
\def\Atop{A^{{\rm top}}_{k+1}}
\def\cpnk{{\rm CP}_{\!n-1,k}^{{\rm top}}}
\def\mskp{\vskip-.25truecm\noindent}
%
%
\voffset=0.35truein\hoffset=0.250truein
\hsize=6.0truein\vsize=8.5 truein
\def\abstr{We review the superconformal properties of 2d matter coupled to
gravity, and extensions thereof. Focusing on topological
strings, we recall how the superconformal structure helps to provide a direct
link between Liouville theory coupled to matter, and matrix models. We also
construct an infinite class of new theories based on $W$-gravity.
}

\font\ninerm=cmr9
\font\ninebf=cmbx9
\font\titsm=cmr10 scaled\magstep2
\nopagenumbers
\def\pubnum{
\hbox{CERN-TH.6812/93}
\hbox{hepth@xxx/9302143}}
\def\pdate{
\hbox{CERN-TH.6812/93}
\hbox{February 1993}
}
\titlepage
\vskip 2.5truecm
\title{\titsm Chiral Rings in Topological (W-)Gravity}
\bigskip
\bigskip
\bigskip
\tenpoint

\font\ninerm=cmr9
\font\ninebf=cmbx9
\centerline{W.\ Lerche}
\bigskip
\centerline{{\it CERN, CH 1211 Geneva 23, Switzerland}}
\bigskip
\vfil
{\centerline{\it Talk given at the }}
{\centerline{\it 26th Workshop: ``From Superstrings to Supergravity"}}
{\centerline{\it Erice - Sicily, 5-12 December 1992}}
\bigskip
\bigskip\vfil
\noindent{\tenrm \abstr}
\vskip 3.truecm
\eject
\def\pdate{}
\centerline{{ \ninebf CHIRAL RINGS IN TOPOLOGICAL (W-)GRAVITY}}
\vskip.8truecm
\centerline{{\ninerm W.\ LERCHE}}
\centerline{{\it CERN, CH 1211 Geneva 23, Switzerland}}
\vskip.5truecm
\vbox{\hbox{\centerline{{\ninerm ABSTRACT}}}
{\smallskip\leftskip 3pc \rightskip 3pc \noindent \ninerm \abstr\smallskip}}

\footline={\hss\tenrm\folio\hss}

\newsec{Introduction}

There has been some recent progress in understanding theories describing 2d
matter coupled to 2d gravity. I would like to review here in particular the
findings of ref.\BLNWB. That paper deals actually with two different and
logically independent lines of development: one is to exploit an \nex2 \sc
symmetry that all 2d matter-gravity systems have. The other is an extension
to $W$-gravity coupled to matter, which adds an infinite sequence of new
theories.

Matter-plus-gravity systems are interesting to study because they are, for
certain choices of matter theories, supposed to be exactly solvable. More
precisely, they are supposed to be equivalent to matrix models\mat, which are
exactly solvable by themselves as a consequence of an underlying structure of
KdV-like integrable hierarchies. To deduce this equivalence directly from
\lv\ theory appears to be difficult, largely due to technical complications. We
will show below that the above-mentioned \nex2 \sc structure provides
a manifest and direct relationship of (at least certain of) such models to
matrix models.

We like to start by briefly recapitulating ordinary gravity coupled to
conformal 2d matter. For simplicity, we will consider mainly minimal
matter models, but this is not really important. These matter models, denoted
by $\mm pq$, where $p,q=1,2,\dots$ are coprime integers, have central
charges
$$
c_M \ =\ 13 - 6 (t + \coeff1t)\ , \qquad\ \ \ t\ \equiv\ q/p\ .\eqn\cmdef
$$
It is well known that such theories can be realized in terms of a single free
boson $\phi_M$, with an appropriate amount of background charge $\a_0$; the
stress tensor has the familiar form: $T_M = -(1/2)(\del\phi_{M})^2 +
i(\a_0/\sqrt2)\del^2\!\phi_M$, where $\a_0 \equiv (1-t)/\sqrt t$. The
primary fields can be represented by vertex operators $\Phi_{r;s} =
V^M_{r;s} \equiv e^{i\a^M_{r;s}\phi_M}$, with $\a^M_{r;s}=\coeff1{\sqrt2}
[\a_+(r-1)+\a_-(s-1)]$, where $\a_+\equiv \sqrt t$, $\a_-\equiv -1/\sqrt t$.
For minimal models $\mm pq$, one restricts the labels to the Kac table, that
is, to $1 \leq r \leq p-1$ and $1\leq s \leq q-1$.

For $t=1$, the matter model is not minimal, but becomes the celebrated $c=1$
theory that received quite some attention due to its relationship with 2d black
holes\blackh. For generic $t$, the central charge $c_M$ is not equal to 26.
Therefore, when coupling to 2d gravity, one deals with ``non-critical'' strings
in which gravity is governed by the \lv\ degree of freedom. \lv\ theory is
usually described in a way very similar to the above matter theory. One treats
the \lv\ field $\phi_L$ like a free field, and tries to incorporate the
cosmological constant term in a perturbative manner (this is possible
only for the ``topological'' theories; see below). One constructs a stress
tensor similar to the one of the matter model: $T_L = -(1/2) (\del\phi_{L})^2 +
(\beta_0/\sqrt2) \del^2\!\phi_L$and chooses the background charge,
$\b_0=(1+t)/\sqrt t$, such that $c_L+c_M =26$. This precisely offsets the
central charge the ghost stress tensor $T_{gh} = - 2 b (\del c) -(\del b) c$,
so that the total central charge vanishes. Here, $b$ and $c$ denote the
fermionic ghosts with spins equal to $2$ and $-1$.

As usual in \brs quantization, the physical states of the combined
matter-gravity system are given by the non-trivial cohomology classes of
a \brs\ operator. This operator looks \mskp
$$
\qbrs\ =\ \oint\coeff{dz}{2\pi i} \jbrs\ , \qquad \ \ \jbrs\ =\ c\,[T_M + T_L +
\shalf T_{gh}] \eqn\brstcur
$$
and is nilpotent for $c_L+c_M =26$. The most prominent physical
states correspond to the tachyon operators\DK:
$$
T_{r,s}\ = \ c\, V^L_{r;-s}\,V^M_{r;s}\ .\eqn\tach
$$
By convention, these operators have $bc$ ghost number equal to one. In
addition, there exist\EXTRA\ extra physical states whose
number and precise structure depends on the specific value of $t$. For unitary
minimal models, where $t=(p+1)/p$, there exist infinitely many of such extra
states for each matter primary, whereas for generic $t$, there exists basically
only one extra sort of states besides the tachyons: these are the operators
with vanishing ghost charge. They form what is called\Witgr\ the ground
ring; we will denote it by $\RG$. It is precisely because these operators have
zero ghost charge (and zero dimension like all physical operators), that the
set of ground ring operators closes into itself under operator products. In
fact, even though this ring is in general infinite, it is finitely generated,
ie., it has two generators by whose action all other ring elements can be
generated\Witgr:
$$
\eqalign{
x\ &=\ \big[ bc -
{t\over\sqrt{2t}}(\del\phi_L-i\del\phi_M)\big]\,V^L_{1,2}V^M_{1,2}\cr
\g\ &=\ \big[ bc -
{1\over\sqrt{2t}}(\del\phi_L+i\del\phi_M)\big]\,V^L_{2,1}V^M_{2,1}\  .}
\eqn\gx
$$
The rest of the operators with non-zero ghost numbers fall into modules of this
ring\Witgr\cc\modul. The structure of $\RG$ (and that of its cousins obtained
by non-trivial pairings of left- and right moving sectors) characterizes to
some extent a given theory. \goodbreak

\newsec{\nex2 \sc structure}

The properties of the ground ring elements remind very much to the typical
features of chiral primary fields in \nex2 \sc theories. The whole point is, of
course, that the matter-gravity-ghost system is essentially nothing but a
(twisted) \nex2 \sc theory. More precisely, it is known\BGS\ that one can
improve the \brs current \brstcur\ by an irrelevant total derivative piece,
$$
G^+ ~=~ \jbrs - \del\Big(\sqrt{\coeff2t}  (c \del\phi_L) + \shalf
(1 - \coeff2t) \del c \Big)\ , \eqn\mggplus
$$
such that $G^+$ together with \mskp
$$
G^- \ =\ b\ ,\qquad
T\ ~=~ T_L + T_M + T_{gh}\ ,\qquad
J\ ~=~ c b  + \sqrt{\coeff2t}\,\del \phi_L\ ,
\eqn\currents
$$
indeed generates the (topologically twisted\TOPALG\cc\EYtop) \nex2 \sc algebra,
$$
\eqalign{
T(z)\shdot T(w)\ &\sim\
{2T(w)\over\zmw 2}
+{\partial T(w)\over (z-w)}\ ,\cr
T(z)\shdot G^\pm(w)\ &\sim\
{\coeff12(3\mp 1) G^\pm(w)\over\zmw 2}+{\partial G^\pm(w)\over (z-w)}\ ,\cr
T(z)\shdot J(w)\ &\sim\ {\coeff 13c^{N=2}\over\zmw 3} +
{J(w)\over\zmw 2}+{\partial J(w)\over (z-w)}\ ,\cr
J(z)\shdot J(w)\ &\sim\ {\coeff 13c^{N=2}\over\zmw 2}\ ,\ \ \ \ \ \
J(z)\shdot G^\pm(w)\ \sim\ \pm {G^\pm(w)\over (z-w)}\ ,\cr
G^+(z)\shdot G^-(w)\ &\sim\
{\coeff 13c^{N=2}\over\zmw 3}+{J(w)\over\zmw 2}+{T(w)+\partial J(w)
\over (z-w)}\ ,\cr
G^\pm(z)\shdot G^\pm(w)\ &\sim\ 0\ ,\cr
}\eqn\tneqtwo
$$
with anomaly \mskp
$$
c^{N=2} ~=~ 3\big(1 - \coeff2t\big) \ . \eqn\minntwo
$$
Upon untwisting, $T\to T-\shalf\del J$, $c^{N=2}$ becomes the central charge of
an ordinary \nex2 algebra. Note that the free-field realization \mggplus,
\currents\ of the \nex2 algebra is different from the well-known one\Muss. This
is however irrelevant, and one can show\BLNWB\ that the above realization can
be obtained by hamiltonian reduction\bo\ from a $SL(2|1)$ WZW model in a way
that is analogous and equivalent to the usual way of deriving a free-field
realization of the \nex2 algebra.

An immediate question is the one about the significance of the \nex2 \sc
symmetry. For general $t$, the mere presence of an \nex2 algebra doesn't
really seem to provide any important new insights. On the other hand, for
integer $t$, which we like to write as $t=2+k$, a lot can be learned: namely
then the anomaly \minntwo\ becomes equal to the central charge of the \nex2
minimal models, $A_{k+2}$: $c^{N=2}={3k\over k+2}$. This is a powerful
statement, since minimal models tend to be easily solved purely by
representation theory. (For general $t=q/p$, the level $k$ becomes fractional,
which corresponds to non-minimal, in general non-unitary \nex2 superconformal
models. From representation theory alone, not much can be said about such
theories in general.)

However, this does not yet imply that the minimal models $\mm1{2+k}$ coupled to
gravity are the same as the topologically twisted \nex2 minimal models,
$\Atop$. A priori, these theories don't
have, in fact, even the same spectra. The spectrum of topological \nex2 models
is well known\EYtop: it is given by the chiral ring\LVW, which is the finite
set of primary chiral fields. For $\Atop$, this is a nilpotent, polynomial ring
generated by one element:
$$
\RN\ =\ {P(x)\over [x^{k+1}\equiv0]}\ =\ \Big\{ 1,x,x^2,\dots,x^k \Big\}\
.\eqn\nring
$$
One can check that powers of the ground ring generator $x$ in \gx\ are indeed
primary chiral fields with respect to the \nex2 currents \mggplus\ and
\currents, and that $\RN$ is identical to the subring of $\RG$ that is
generated by $x$. (It turns out that the corresponding tachyons \tach\ have the
same \nex2 quantum numbers as the ground ring elements, so that they can be
viewed as different representatives\foot{These representatives just happen not
to close among themselves under operator products.} of the same set of primary
chiral fields. This is not true for general $t$.)

However, the ground ring $\RG$ of the matter-gravity system contains infinitely
many more operators\Ind:
$$
\RG\ =\ \RN \otimes \Big\{\,(\g)^n, \ \ n=0,1,2,\dots\,\Big\}\ .\eqn\gring
$$
These extra operators simply do not exist in the topological minimal models;
they are not primary with respect to the \nex2 algebra.

The difference between the spectra \nring\ and \gring\ can be accounted for as
follows: it turns out that these extra operators are exact with respect
to an additional \brs like operator, $\qt$:
$$
\g =\ -\big\{\qt\,,\,\coeff{t+1}t\,\del c
+\coeff1{\sqrt{2t}}\,c\,\del\phi_L\,\big\}\  ,\quad {\rm where}\ \
\qt\ =\
 \oint\coeff{dz}{2\pi i}\,b\,e^{-\coeff t{\sqrt{2t}}(\phi_L-i\phi_M)}\ .
\eqn\qtilde
$$
One can show\BLNWB\ that $\qt$ is one of the Felder-like screening operators
that arise in our particular free field realization of the minimal models. That
is, {\it by definition} the full \brs operator of the topologically twisted
\nex2 minimal models is the sum of $\qbrs$, $\qt$ and other screening
operators, and it truncates the infinite free field spectrum precisely to the
finite set of physical operators \nring.

We thus see that this full \brs operator of the topological minimal models is
not the correct one if we wish to describe the minimal models $\mm1{2+k}$
coupled to gravity. The correct operator obtains if we drop $\qt$ as an extra
\brs operator; it can be shown that then the ``missing'' operators $(\g)^n$
become physical. There is actually a better way to formulate this in terms of
equivariant cohomology\topgr, but for lack of space we refrain from doing so.

It can be shown\EYQ\ that these {\it modified} minimal topological models,
which contain the operators $(\g)^n$ and which are equivalent to the minimal
models $\mm1{2+k}$ coupled to gravity, are in fact also equivalent to the
models $\Atop$ coupled to {\it topological} gravity\topgr. A priori, the
building blocks of $\Atop$ and of the same models coupled\Keke\ to topological
gravity appear to be quite different. There is, however, the remarkable fact
that the total \brs operator of the topological matter plus topological gravity
system obeys\EYQ
$$
\cQ^{tot}\ \equiv \cQ^{N=2} + \qbrs  \ =\ U^{-1}\cQ^{N=2} U \ ,
\eqn\qequiv
$$
where $U$ is some homotopy operator. The upshot is that the cohomologies
of $\Atop$ coupled to topological gravity and of the modified minimal
topological models are isomorphic, so that at least at the level of Fock spaces
$$
\eqalign{
\Big[\,\mm1{2+k}\otimes{\rm \lv\ gravity}\,\Big] \ &\cong\
\Big[\,\Atop\attac{{{\rm modified}\atop{\rm cohomology}}}\!\!\!\!\!\Big]\cr
&\cong\ \Big[\,\Atop\otimes{\rm topological\ gravity}\,\Big]\ .}
\eqn\equivs
$$
In view of this equivalence, the extra ground ring elements, $(\g)^n$, can thus
be interpreted as topological gravitational descendants, and the full ground
ring \gring\ might be called a ``gravitationally extended chiral ring'' of
the topological matter model. Note also that for $k=0$, which corresponds to
the trivial topological matter theory, the LHS of \equivs\ turns precisely into
Distler's formulation of topological gravity\dis\cc\BMPtop.

\noindent With \equivs\ at hand, it is then easy to conclude that
$$
\Big[\,\mm1{2+k}\otimes{\rm \lv\ gravity}\,\Big] \ \cong\
\Big[\,{\rm matrix\ model\ of\ type\ }(1,2+k)\,\Big]\ ,
\eqn\matmod
$$
as it is supposed to be the case\MD. This is a direct consequence of the
fact\Keke\ that the recursion relations of $[\Atop\otimes{\rm topological\
gravity}]$ are the same as those of the matrix models\TFTmat.

One would like to make similar statements for more general models
$\mm pq$ coupled to gravity; such theories can be considered as deformations of
the above, topological ones, and it is an interesting question whether
\nex2 language would be useful for describing this.

The equivalence \matmod\ can be exhibited also in a more direct way. The point
is that the \LG formulation\Vafa\ of the topological matter models, $\Atop$,
can be directly related\DVV\ to the KdV structure of the matrix models. More
precisely, the \LG superpotential, which describes the effect of
perturbations\foot{Note that such perturbations lead away from the conformal
point. We restrict here to the ``small phase space'', ie., to perturbations
generated by the primary fields.}$\exp\int d^2\!z\,d^2\!\theta\sum_{i=0}^k t_i
x^i$, has the generic form \mskp
$$
W(x,t_i)\ =\ \coeff1{k+2}\,x^{k+2} - \sum_{i=0}^k g_i(t_j)x^i\ ,
\eqn\suppot
$$ \mskp
where the coupling constants $g_i(t_j)$ are certain, in general non-trivial
functions of the perturbation parameters. Since the correlation functions can
easily be computed\DVV\ from $W(x,t_i)$, solving the theory amounts to
determining these functions. They can be obtained by solving the differential
equations
$$
{\del\over\del t_i}W(x,t)\ =\ \coeff1{i+1}((k+2)W)^{{i+1\over k+2}}_+\ ,
\eqn\wdeq
$$
which just express the fact that the $t_i$ are ``flat'' coordinates of the
deformation space. The crucial observation\DVV\ is that under the substitutions
$x\to D\equiv {\del\over\del z}$ and $W(x,t_i)\to L(D,{t_{i+1}\over i+1})$,
these equations becomes precisely the dispersionless, quasi-classical
limit\disp\ of the KdV flow equations
$$
{\del\over\del t_i}L(D,t) \ =\ \big[\,(L^{i\over k+2})_+,L\,\big]\ .
\eqn\kdv
$$
These equations describe\mat\cc\MD\ (or even define) the dynamics of the matrix
models of type $(1,k+2)$. This immediately proves the equality of (genus zero)
correlation functions of the primary fields of $\Atop$ with the corresponding
correlators of the matrix models. These arguments, which involve only \nex2 \LG
theory, can in fact be extended to gravitational descendants and to
recursion relations they obey\Loss\cc\EYQ. This is precisely in the spirit of
what was said above: the ingredients of the coupling of $\Atop$ to topological
gravity are already build in the structure of the models $\Atop$ themselves;
all what is necessary to describe the coupling of these models to topological
gravity is to modify their cohomological definition.

Of particular interest is the perturbation of these models by the
``cosmological constant'' term\foot{We define this here as being the operator
with trivial matter piece. For the topological models the dependence on $\mu$
is analytic and thus perturbation theory around $\mu=0$ is perfectly well
defined. This is in contrast to general models $\mm pq$ coupled to gravity.}.
In our language, it is the perturbation by the top element of $\RN$,
$$
S_{{\rm cosm}}=\mu \int d^2\!z\, e^{\sqrt{{2\over t}} \phi_L}\ =\ t_k \int
d^2\!z\,d^2\!\theta\, x^k\ .
\eqn\cosmo
$$
It is known\integr\ that this perturbation is integrable and leads to the
massive quantum \nex2 sine-Gordon model; although not invariant under the full
(twisted) \nex2\goodbreak\noindent \sc symmetry, it is supersymmetric, and the
(corrected) supercharge $\oint G^+$ still serves as a \brs operator. The
effective superpotential is given by a Chebyshev polynomial:
$$
W(x,t_k=\mu)\ =\
\coeff2{k+2}\mu^{{k+2}\over2}\,T_{k+2}(\shalf\mu^{-{1\over2}}x)\ =\
\coeff1{k+2}\,x^{k+2}-\mu\, x^k + O(\mu^2)\ .
\eqn\cheby
$$
At $\mu=1$, the deformed chiral ring becomes identical\fusions\ to
the fusion ring of the $SU(2)_k$ WZW model, which it is also the same as the
operator product algebra of the $SU(2)_k/SU(2)_k$ topological field theory.
This observation then allows to finally make contact to the formulation of
matter-plus-gravity models in terms of topological $G/G$ theories\GG:
it is known\GG\cc\MuVa\ that at the level of Fock space cohomology,
the (suitably defined) $SU(2)/SU(2)$ model is indeed equivalent to the
matter-gravity system. We thus have in addition:
$$
\Big[\,\mm1{2+k}\otimes{\rm \lv\ gravity}\,\Big]\attac{\mu=1}\!\!\cong\ \
\Big[\,{SU(2)_k\over SU(2)_k}\attac{{{\rm modified}\atop{\rm
cohomology}}}\!\!\!\!\!\Big]
\eqn\otherequivs
$$

\newsec{Generalizations}

The other line of development taken up in ref.\BLNWB\ is the generalization
to $W$-gravity\Wrev\ coupled to $W$-matter. Here one considers tensor products
$$
 W_n^{\rm matter}\otimes W_n^{\rm \lv}\otimes_{j=1}^{n-1}
\{b_j, c_j\} \ , \eqn\wstring
$$
which might be called ``non-critical $W$-strings''.\Wstrings
Above, $W_n^{\rm matter}$ denotes conformal field theories that have a
$W$-algebra as their maximal chiral algebra, which can be for example $W_n$
minimal models $\wnpq npq$ with central charges $c_M = (n-1)[ 1-n (n+1)
{(t-1)^2\over t}],\ t=q/p$. Furthermore, $W_n^{\rm \lv}$ denotes a
$(n-1)$-component generalization of \lv\ theory (Toda theory), and $\{b_j,
c_j\}$ denotes the Hilbert space of a ghost system with spins $j+1$ and $-j$,
respectively. As it turns out, the structure of these theories for arbitrary
$n$ is very similar to $n=2$, which corresponds to ordinary gravity. However,
only for $n=3$ the generalization\foot{The existence of \brs currents for
arbitrary $n$ can be inferred from indirect arguments\BLNWB.} of the \brs
current is explicitly known\wbrs:
$$
\eqalign{
\jbrs\ &=\ \ci2 \big[\coeff1{b_{L}}W_L  + \coeff i{b_{M}} W_M \big]
+ \ci1 \big[T_L +T_M +\shalf\tg^{1} +\tg^{2} \big]\cr
&+ \big[T_L -T_M \big]\bi1 \ci2(\del\ci2)
+ \mu(\del\bi1)\ci2(\del^2\!\ci2)+ \nu\bi1 \ci2(\del^3\!\ci2)\ ,}\eqn\jb
$$
where $b^2_{L,M}\equiv{16\over5c_{L,M}+22}$ and $\mu={3 \over 5}\nu= {1 \over
10{b_L}^2} (1-17{b_L}^2)$. In this equation, $T_{L,M}$ and $W_{L,M}$ denote the
usual stress tensors and $W$-generators of the \lv\ and matter sectors, and
$\tg^i$ are the stress tensors of the ghosts.

Using this \brs current, one can study the spectrum of physical operators of
$W_3$ matter coupled to $W_3$ gravity, and finds\BLNWB\cc\popextra\cc\newBMP\
that the analogs of ground ring elements and tachyons are states with ghost
numbers equal to $0,1,2,3$ (the first number corresponds to ground ring
elements, and the last one to tachyons). The \goodbreak\noindent explicit
expressions are however too complicated to be written down here.

The interesting point is that there appears an \nex2 \sc
symmetry for all $n$. For example, for $W_3$ gravity one finds that
$$
\eqalign{ G^+\ &=\ \jbrs +
\del\Big[\,
-c_{1}J+2i\sqrt{\coeff t3} b_{1} c_{1} c_{2}J +
  i\coeff{\left( 1 + t \right) }{2} \sqrt{\coeff3t}
   b_{1} c_{1} (\del c_{2})
   \cr&-
    i\coeff{\left( 3 + 2 t \right) }{\sqrt{3 t}}
   b_{1} (\del c_{1}) c_{2} -
    \coeff{(7 {t^2}-10t-15)}{4 t}   b_{1} (\del^{2}\!c_{2}) c_{2}+
  i\coeff{(t-9)}{\sqrt{3 t}} b_{2} (\del c_{2}) c_{2}
  \cr&-
  i\coeff{\left( 3 + 4 t \right) }{\sqrt{3 t}} (\del b_{1}) c_{1} c_{2}-
  \coeff{3 (4 {t^2}-2t-3)}{2 t} (\del b_{1}) (\del c_{2}) c_{2} +
 \coeff{(t-3)}{t} (\del c_{1})
  \cr&+
 i\coeff{1}{2 \sqrt{3 t}} c_{2}[2t {J^2}-3 (t-5)T_L-3 (t-1)T_M
 -6( 1 + t )\del J]
  \cr&+
  i\coeff{ \left( 1 + t \right) }{2} \sqrt{\coeff3t}(\del c_{2})J-
  i\coeff{({t^2}-4t-1)}{2t}\sqrt{\coeff3t} (\del^{2}\!c_{2}) +
  t b_{1} (\del c_{2}) c_{2}J
  \, \Big]\ ,}\eqn\wbrs
 $$
together with \mskp
$$
\eqalign{
G^- \ &=\ b_1\ ,\qquad\ \ \ \
T\ ~=~ T_L + T_M + T_{gh} \ ,\cr
J\ ~&=~ c_{1} b_{1}  + c_{2} b_{2}  +
  \coeff3{\sqrt t}(\l_1\cdot \del \phi_L)
 +\coeff i2\sqrt{\coeff3t}  (t-1) \del[b_{1}c_{2}] \cr
}\eqn\wcurrents
$$
gives a non-standard free field realization of the topological algebra
\tneqtwo\ with
$$
c^{N=2} ~=~ 6\big(1 - \coeff3t\big) \ . \eqn\cks
$$

Since we are dealing here with theories with an extended symmetry, coupled to
an extended ``$W$-geometry'', it is perhaps not too surprising to find that
these topological algebras actually extend to topologically twisted \nex2
$W$-algebras. For $t=n+k$, which corresponds to $W_n$-minimal matter models
$\wnpq n1{n+k}$, the anomaly indeed becomes equal to the central charges of the
minimal \nex2 $W_n$ models at level $k$: $c^{N=2}=3{(n-1)k\over n+k}$. These
models are just the well-known Kazama-Suzuki models\KS\ based on cosets
$SU(n)_k\over U(n-1)$, which are known to have an \nex2 $W_n$ chiral
algebra\Wchiral. The models that arise here are of course the
topologically twisted versions, which we will denote by $\cpnk$.

The chiral rings of these topological $W_n$ matter models are well
understood\LVW\cc\cring. They are generated by primary chiral fields
$x_i$, $i=1,\dots,n-1$ (with $U(1)$ charges equal to $i/(n+k)$), and
have elements \mskp
$$
\cR^{\cpnk}\ =\ \Big\{\prod_{i=1}^{n-1}(x_i)^{m_i}\ ,\sum n_i\leq k\,\Big\}\ .
\eqn\ksrings
$$ \mskp
On the other hand, the full ground rings of the minimal models $\wnpq n1{n+k}$
coupled to $W_n$-gravity contain in addition generators $\g_i$, $i=1,\dots,n-1$
(with $U(1)$ charges equal to $i$) and appear to be ``$W$-gravitationally
extended'' chiral rings of the Kazama-Suzuki models: \mskp
$$
\RG\ =\ \cR^{\cpnk} \otimes \Big\{\prod_{i=1}^{n-1}(\g_i)^{n_i}\ ,\
n_i=1,2,\dots\,\Big\}\ .\eqn\fullring
$$ \goodbreak\noindent\mskp
Although this has not yet been thoroughly investigated for general $n$, our
considerations seem so far to indicate that the structure is
completely analogous to the one of $n=2$. Accordingly, one would have for
topological $W$-strings:
$$
\Big[\,\wnpq n1{n+k}\otimes W_n{\rm\hyp gravity}\,\Big]
\ \cong\ \Big[\,\cpnk\otimes{\rm topological}\,\ W_n{\rm\hyp gravity}\,\Big]\ ,
\eqn\Wequivs
$$
as well as the obvious generalization of \otherequivs. However, the building
blocks of topological $W$-gravity as constructed in ref.\topw\ look different,
and the precise connection of our considerations with those of ref.\topw\ needs
to be investigated in more detail.

\newsec{Concluding remarks}

One possible motivation for investigating the above kind of generalizations is
the wish to step beyond the $c_M\!=\!1$ barrier of ordinary gravity. For a
given $W_n$-gravity, there is a barrier at $c_M\!=\!n\!-\!1$, below of which
there is a finite number of (dressed) primary fields and below of which the
theory should be solvable. In analogy to ordinary gravity, one would even
expect that such theories should be solvable also at the accumulation points
$c_M\!=\!n\!-\!1$ (where there exists an extra $SU(n)$ current algebra
symmetry). At these points, such models are presumably related to black
hole type of objects in spacetimes with signature $(\!n\!-\!1,\!n\!-\!1)$ and
are characterized by topological field theories based on non-compact versions
of $\cpnk$.

Another motivation would be to find analogs of the identification \matmod. At
the moment it is not clear to us whether any such generalized matrix models
really exist, but at least the corresponding generalizations of the KdV
hierarchy appear to exist; details will be presented elsewhere.

A further point of view might be that there is in fact nothing special about
$W$-extensions of minimal models: one is dealing with just particular examples
of extended chiral algebras, $\cA$. One is tempted to speculate that the
emergence of an \nex2 structure, of extensions of topological gravity and
finally of generalized integrable hierarchies might be a more general
phenomenon, and occurs even for arbitrary RCFT (or at least, for a large class
thereof). The conjecture that for an arbitrary chiral algebra $\cA$, a theory
of the form $$ \Big[\,\cA{\rm \hyp minimal\ matter}\otimes``\!\cA{\rm \hyp
gravity"}\,\Big] $$\mskp might be equivalent to some ``topological $\cA$ string
theory'', seems to be related to the recent ideas of Gepner. A conjecture put
forward in ref.\gep\ is that there is an \nex2 \sc field theory associated with
{\it any} RCFT, the chiral ring of the \nex2 theory being isomorphic to the
fusion ring of the RCFT. This association is very similar to what ``coupling to
$\cA$-gravity'' would achieve: gauging the chiral algebra $\cA$ implies that
all $\cA$ descendants become unphysical (presumably by becoming top components
of \nex2 supermultiplets), and the spectrum truncates to the (dressed) primary
fields (apart from possible gravitational descendants). The algebra of these
fields is then more or less the chiral ring of ref.\gep.\goodbreak

\newsec{Acknowledgements}
\vskip-.2truecm
The work described in this lecture is based on an enjoyable collaboration with
M.\ Bershadsky, D.\ Nemeschansky and N.P.\ Warner. I am also thankful to Keke
Li for his challenging criticism two years ago. In addition I would like to
thank the organizers of the workshop for their hard work, and for providing an
opportunity for me to present this material.
\vskip.2truecm
\refout
\end